\documentclass{mn2e}

\usepackage{graphics,amssymb}

\title[The 1319~\AA\ emission line]
{The emission line near 1319~\AA\ in solar and stellar spectra}
\author[C. Jordan]
{C.Jordan$^1$\thanks{E-mail: cj@thphys.ox.ac.uk} \\
$^1$Department of Physics, The Rudolph Peierls Centre for Theoretical 
Physics, University of Oxford, 1 Keble Road, Oxford, OX1\,3NP}
\begin{document}

\date{Accepted 2011 January 25. Received 2011 January 24; in original form 
2010 September 30}

\pagerange{\pageref{firstpage}--\pageref{lastpage}} \pubyear{2010}

\maketitle

\label{firstpage}

\begin{abstract}
An emission line at $\simeq 1319$~\AA\ is one of the strongest unidentified
 lines in the ultraviolet spectra of cool dwarf stars. In most lists of 
solar and stellar lines it is identified as a transition in N\,{\sc i}, 
although its intensity would then be anomalous and the wavelength does not 
precisely fit that expected for N\,{\sc i}. The line is also observed in giant
 stars but becomes very weak in supergiants, relative to photoexcited lines 
of neutral atoms. The measured wavelength of the line in stellar spectra is 
1318.94~$\pm 0.01$~\AA. Observations of giant stars provide further 
information that shows that this line is {\it not} due to N\,{\sc i}. 
It is proposed that the line is due to a decay from the 
3p$^3$($^2$D$^{\rm o})$3d~$^1$D$^{\rm o}_2$ level of S\,{\sc i}, above the 
first ionization limit. The previous tentative assignment of this upper level 
to a S\,{\sc i} line at $\simeq 1309.3$~\AA\ then needs to be revised. The 
1309.3-\AA\ line is identified here for the first time in an astrophysical 
source. The 3d~$^1$D$^{\rm o}_2$ level could, in principle, be populated by 
collisions from nearby autoionizing levels that are shown to have large number 
densities, through population by low-temperature dielectronic capture. 
Spin-orbit interaction with the autoionizing 3d~$^3$D$^{\rm o}_2$ 
level might also lead to dielectronic capture into the 3d~$^1$D$^{\rm o}_2$ 
level. A line at 1309.87~\AA\ observed in cool giant stars is identified as a 
transition in P\,{\sc ii}, pumped by the O\,{\sc i} resonance lines.  
\end{abstract}


\section{Introduction}
Some years ago, attention was drawn to the unsatisfactory identification of a 
line near 1319~\AA\ with the close N\,{\sc i} blend at 1319.001~\AA, which 
arises from the 2p$^3$~$^2$P$^{\rm o}_{1/2,3/2}$ -- 2p$^2$3d~$^2$P$_{1/2}$ 
transitions (Jordan \& Judge 1984; Jordan 1988). It was hoped 
that further laboratory or theoretical studies of neutral atoms would shed 
light on this problem. There are two difficulties with the identification 
with N\,{\sc i}; the other closely blended members of the multiplet (the 
$^2$P$^{\rm o}_{1/2. 3/2}$ -- $^2$P$_{3/2}$ transitions at 1319.675~\AA) are 
very much weaker than expected, and the observed wavelength does not 
precisely match that of the N\,{\sc i} transition at 1319.001~\AA.
According to the oscillator strengths given by Kurucz \& Peytremann
(1975), the blend at 1319.675~\AA\ should be a factor of 1.6 stronger than
that at 1319.000~\AA; in L-S coupling, this ratio would be 2.0. The observed
ratio in the Sun is $\le 0.23$ (see Section 2.2). 

Burton \& Ridgeley (1970) first noted the presence of an unidentified line at 
1318.95~\AA\ and suggested that it was probably due to a neutral atom, 
owing to its behaviour at the solar limb. Chipman \& Bruner (1975) 
measured a wavelength of 1318.92~\AA, using their high resolution
 stigmatic spectra. Thus there were early indications that the line near
1319~\AA\ is not due to N\,{\sc i}. 

It is timely to revisit this problem, since the line near 1319~\AA\ is still
being identified with the N\,{\sc i} transition, with the danger that it
could be adopted as a secondary wavelength standard. For example, the line 
appears as N\,{\sc i} in the SUMER (Solar Ultraviolet Measurements of Emitted 
Radiation) spectral atlas (Curdt et al. 2001) and in high resolution stellar 
spectra obtained with the Space Telescope Imaging Spectrograph (STIS) by
 Ayres, Hodges-Kluck \& Brown (2007). In dwarf stars, the line at 
$\simeq 1319$~\AA\ is the strongest in the region between 1240 to 1400~\AA\ 
for which there has not been a satisfactory identification and it deserves 
further discussion.

Although the early observations used were made with the {\it International 
Ultraviolet Explorer} ({\it IUE}), the higher sensitivity and spectral 
resolution of the Goddard High Resolution Spectrograph (GHRS) and the STIS on 
the {\it Hubble Space Telescope} ({\it HST}) have led to     
greatly improved stellar spectra. Thus it is now possible to study the line at
 $\simeq$ 1319~\AA\ in other main-sequence stars and in several giant stars, 
in the context of other lines observed.

Similarly, the SUMER instrument on {\it SOHO} ({\it Solar and Heliospheric 
Observatory}) has provided high spectral and spatial resolution observations 
of the solar far-ultraviolet ({\it fuv}) and {\it uv} spectrum between 465 
and 1610~\AA, with an improved intensity calibration. The early photographic 
spectra obtained with the Naval Research Laboratory's (NRL) High Resolution 
Spectrograph and Telescope (HRST) provide limb-to-disc stigmatic observations 
that are still valuable in line identification work [see e.g. figs 1, 2 
(plates 14, 15) in Jordan et al. 1978].   

In Section 2 the observed properties of the line at $\simeq 1319$~\AA\ are 
summarized. Section 3 sets out the possible origins of this line, as a 
transition in S\,{\sc i}. The mechanisms by which the upper levels in 
S\,{\sc i} are populated are discussed in Section 4. In Section 5, a line that
 appears near 1310~\AA\ in the spectra of giant stars is identified 
with a transition in P\,{\sc ii}. The conclusions are summarized in Section 6. 

\section{The observed properties of the 1319-\AA\ line}

\subsection{The wavelength} 
There is a small difference between the measured wavelength of the 
observed line and that expected for the (blended) N\,{\sc i} transition. The 
latter should have a mean wavelength of 1319.001~\AA\ (Moore 1975), assuming  
intensity contributions according to LS-coupling, or adopting the oscillator
strengths of Kurucz \& Petreyman (1975).

From solar spectra, the wavelengths measured by Sandlin et al. (1986) and 
Curdt et al. (2001) are 1318.99~$\pm 0.01$~\AA\ and 1318.98~$\pm 0.01$~\AA, 
respectively. Neither measurement is formally consistent with the wavelength 
and uncertainty reported here from the stellar spectra 
(1318.94~$\pm 0.01$~\AA, see below). Chipman \& Bruner (1975) gave a
 shorter wavelength, 1318.92~\AA, and also claim an overall wavelength 
accuracy of $\pm 0.01$~\AA. The laboratory wavelengths of lines that they 
identified in the vicinity of 1319~\AA\ agree with their measured values to 
within $\pm 0.02$~\AA.  With this degree of accuracy, the wavelength that they
measure is consistent with the value derived below. In a solar flare spectrum 
observed from {\it Skylab} (Cohen, Feldman \& Doschek 1978), the wavelength is
 reported as 1318.95~\AA, with a typical accuracy of $\pm 0.01$~\AA\ for narrow
 lines. Again this is consistent with the present value. The N\,{\sc i} line 
at 1319.675~\AA\ was also observed in the above flare spectrum at its expected
 wavelength, but was a factor of 3.8 weaker than the 1318.95-\AA\ line. 

The line near 1319~\AA\ has now been observed in a wide range of main-sequence
and giant stars. Here we concentrate on observations made with the STIS
instrument on the {\it HST}. In addition to our own spectrum of $\epsilon$~Eri
(HD 22049), the spectra used have been obtained from the archive at the Space 
Telescope Science Institute. For $\xi$~Boo~A (HD 131156A) and $\epsilon$~Eri, 
wavelengths were measured by H. Kay and S. Sim, respectively, as part of 
our wider studies of these stars (private communications). Gaussian profiles
were fitted to the lines. The wavelength of the line near 1319~\AA\ was then
found relative to nearby identified lines of S\,{\sc i} (1300.907~\AA) and 
C\,{\sc i} (1311.363~\AA). For all the stars, the relative wavelengths were
measured using eye fits to the line centres ($\xi$~Boo~A and $\epsilon$~Eri 
were included to test that this was adequate). The results are summarized in 
Table 1. The mean wavelengths measured are 1318.946~\AA\ and 1318.942~\AA, 
with an overall mean value of 1318.94~$\pm 0.01$~\AA. The uncertainty given 
reflects the differences between individual measurements. 

The N\,{\sc i} line at 1319.675~\AA\ also appears near the solar limb and in 
active regions (see e.g. fig. 5 in Jordan 1988), but is much weaker than the 
line at 1318.94~\AA. It is also present in the spectra of cool giants with 
enhanced N/C relative abundances, e.g in 24~UMa (HD 82210), $\mu$~Vel (HD 
92497), $\iota$~Cap (HD 203387) and $\beta$~Cet (HD 4128). Using the known 
interval between the N\,{\sc i} lines, the weaker member of the N\,{\sc i} 
multiplet at 1319.001~\AA\ then lies in the red wing of the stronger line at 
1318.94~\AA, but should have little effect on the wavelengths measured for 
this line. 

Fig. 1 shows the region 1318.6 - 1320.0~\AA\ in two spectra of $\beta$~Cet, 
obtained with the STIS. Note that the STIS wavelength calibration places the 
1318.94~\AA\ line at 1319.00~\AA\ (line 'a'), in disagreement with that 
derived from the measured relative wavelengths given in Table 1. The only line
 present that appears in {\it both} spectra and could be the stronger member 
of the N\,{\sc i} multiplet is that at $\simeq$ 1319.74~\AA\ (line 'c'), about
 0.065~\AA\ longer than expected. Then the weaker member of the N\,{\sc i} 
multiplet would lie at $\simeq$ 1319.07~\AA\ (marked 'b'). 

\begin{table}
\caption{The wavelength of the observed line in late-type stars}
\begin{tabular}{lccc}
\hline \\
Star           &  Type &$\lambda^a$ &$\lambda^b$   \\
               &       &   (\AA)        &  (\AA)               \\
\hline \\
HD 20630   & G5~V  &  1318.952      &  1318.944            \\
HD 131156A    & G8~V  &  1318.946      &  1318.940            \\
HD 10700      & G8~V  &  1318.944      &  1318.945            \\ 
HD 22049  & K2~V  &  1318.944      &     -~$^c$           \\
HD 82210         & G4~III/IV& 1318.947    &  1318.934            \\
HD 92497  & G5~III+dF& 1318.942    &  1318.940            \\
HD 203387     & G8~III&  1318.949      &  1318.940            \\
HD 4128     & K0~III&  1318.949      &  1318.952            \\
HD 124897    & K1~III&  1318.939      &  1318.937            \\
               & Mean  &   1318.946($\pm 0.01$)& 1318.942($\pm 0.01$) \\
\hline \\
\end{tabular}
$^a$ Measured relative to S\,{\sc i} 1300.907~\AA. 
$^b$ Measured relative to C\,{\sc i} 1311.363~\AA. $^c$ A chip problem 
prevents a measurement. \\
\end{table}

\begin{figure}
\resizebox{\hsize}{!}{\includegraphics{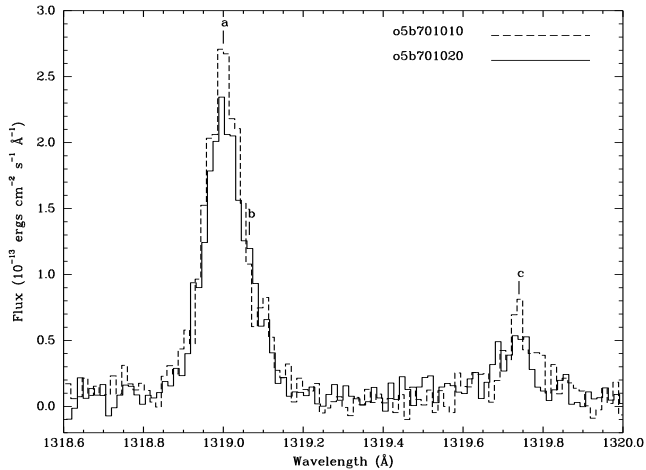}}
\caption{Two STIS spectra of $\beta$~Cet. The line measured here as
occuring at 1318.94~\AA\ is put at 1319.00~\AA\ by the local STIS calibration.
If line 'c' is identified with the theoretically stronger N\,{\sc i} line at 
1319.675~\AA, then the weaker N\,{\sc i} at 1319.001~\AA\ lies in the red wing
 of the line at 1318.94~\AA\ (line 'a') at 1319.07~\AA\ (marked 'b').}
\end{figure}

\subsection{The intensity and spatial distribution in the Sun} 
The spatial distribution of the various emission lines shown in figs 1 -- 4
(plates 14 -- 17) in Jordan et al. (1978) is very valuable in identifying the
lines present. A 20~s exposure obtained from the HRTS-I flight is shown in 
Fig.~2. (This spectrum was provided by the late G. Brueckner and is shown by
permission of G. Doschek, current Head of Solar Physics at NRL.) 
The overall spatial distribution
 of the line at 1318.94~\AA\ (marked 'a') is very similar to that of the 
S\,{\sc i} line at 1300.907~\AA\ (marked 'w'), a property first noted by
Chipman \& Bruner (1975). The line marked 'z' is the C\,{\sc i} line at 
1311.363~\AA. Other lines in Fig.~2 are discussed in Section 3. 

\begin{figure}
\resizebox{\hsize}{!}{\includegraphics{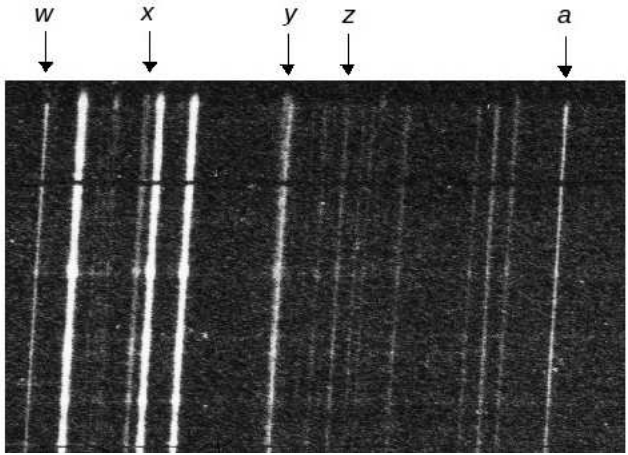}}
\caption{A 20~s exposure of the region from 1300 to 1320~\AA, obtained during
the HRTS-I rocket flight, showing lines of S\,{\sc i} (w), Si\,{\sc ii} (x and
y), and C\,{\sc i} (z) at 1300.907, 1304.370. 1309.276 and 1311.363~\AA, 
respectively. The line at 1318.94~\AA\ is also included (a). A sharp line at
1309.36 $\pm 0.02$~\AA\ appears in the red wing of the Si\,{\sc ii} line (y).
The N\,{\sc i} line at 1319.675~\AA\ is barely visible. Shown by permission of
G. Doschek (NRL).}
\end{figure}

The spatial distribution of the line at 1318.94~\AA\ shows that it is a 
transition in a neutral atom. The energy level schemes of C\,{\sc i}, 
N\,{\sc i}, O\,{\sc i}, Si\,{\sc i} and S\,{\sc i} have been examined for
possible transitions. Most of the suitable unknown energy levels in these 
atoms occur above the first ionization limit. There are observed examples of
narrow decays from theoretically non-autoionizing levels in C\,{\sc i} 
(around 1329~\AA), Si\,{\sc i} (around 1256~\AA) and in S\,{\sc i} at 
1300.907~\AA.

Sandlin et al. (1986) give the peak intensities of solar lines between 
1175~\AA\ and 1680~\AA\ in the quiet Sun (QS), a plage region (P) and a 
sunspot umbral region (U). These can be used to investigate the behaviour of 
lines of interest since there is little difference between the widths of 
lines of neutral atoms (0.08~\AA\ for Si\,{\sc i}, 0.085~\AA\ for O\,{\sc i} 
and S\,{\sc i}, and 0.09~\AA\ for C\,{\sc i} and N\,{\sc i}). The {\it known}
 lines of these elements have characteristic orderings of the intensity ratios
I(P)/I(QS), I(U)/I(QS) and I(P)/I(U), which ultimately depend on the line 
formation processes and the structure of the solar atmosphere.   
Only lines with peak intensities
of $\ge 100$~erg~cm$^2$~s$^{-1}$~st$^{-1}$~\AA$^{-1}$ are included, except
for the N\,{\sc i} at 1319.675~\AA\ that is of particular interest. 

The lines of S\,{\sc i} discussed below are all transitions to terms of the 
ground state configuration (3s$^2$3p$^4$). Such lines, including that at 
1300.907~\AA, which originates from the 
3p$^3$($^2$D$^{\rm o}$)5s~$^1$D$^{\rm o}_2$ level above the first ionization 
limit, are characterised by intensities for which the P/QS ratios are largest 
($\simeq$~3) and the U/QS ratios are smallest ($\simeq$~1). The average P/U 
intensity ratio is somewhat smaller than the average P/QS intensity ratio. 
The line at 1318.94~\AA\ fits this pattern. 

The lines of C\,{\sc i}, including those near 1329~\AA\ that originate from
above the first ionization limit, show a different pattern; the P/QS ratios
are largest ($\simeq 4.5$) and the U/QS ratios are twice the P/U ratios of 
1.5. 
 
Although the ratios for N\,{\sc i} line at 1319.675~\AA\ have the same order 
as those  of S\,{\sc i}, both the P/QS and P/U ratios are much larger 
($\simeq$ 9 and 7, respectively). 

The intercombination lines of O\,{\sc i} at 1355~\AA\ and 1358~\AA\
are the only (nearly) optically thin lines available. These have the same 
ordering as the lines of S\,{\sc i}, 
but instead of similar P/QS and P/U values they have similar U/QS and P/U 
ratios.

Most lines of Si\,{\sc i} have umbral peak intensities that are too small to 
use. The mean P/QS ratio is much larger ($\simeq$ 7) than for S\,{\sc i}.
The only unblended line (1255.27~\AA) from above the first ionization limit
 has ratios that are most similar to those of N\,{\sc i}.  

Thus the known lines of C\,{\sc i}, N\,{\sc i}, O\,{\sc i} and Si\,{\sc i} do 
{\it not} follow the same pattern as that of the S\,{\sc i} lines and the
line at 1318.94~\AA. 

Since it is likely that the line at 1318.94~\AA\ is due to a transition in 
S\,{\sc i}, the ratio of the peak intensity of this line to that of the 
1300.907-\AA\ line is now examined. This ratio varies
little between different solar features. From Sandlin et al. (1986), assuming 
that the line widths are the same, the ratio is 0.95 in a plage region, 0.80 
in the sunspot umbra and 0.81 in the quiet region. The ratio of peak radiances 
given by Curdt et al. (2001) (in table 1a) are 0.92 (quiet Sun), 1.01 
(sunspot) and 0.96 (coronal hole). The background has not been removed, but in
each case is at least an order of magnitude lower than the peak intensity. 
Over this small wavelength range, the relative radiances measured by Curdt et 
al. (2001) should be more accurate than those from the photographic HRTS 
spectra. Even in the flare recorded by Cohen et al. (1978), the ratio is 0.98, 
similar to that in the quiet Sun. Thus a common origin of the two lines is 
likely.

In contrast, the ratio of the peak intensity of the line at 1319.675~\AA\ to 
that at 1318.94~\AA\ is 0.23 (P), 0.09 (U) and 0.08 (QS), far from the ratio
of 2:1 (LS coupling) to 1.6:1 (Kurucz \& Petreymann 1975) expected if both 
lines were due to N\,{\sc i}. 
   
Curdt et al. (2001) also show the ratio of the mean radiance (as a function of
 wavelength) from regions of supergranulation cell boundaries to those from  
cell interiors (see their fig. 4). They remark that the variation of this 
ratio depends on the
type of emission line (e.g. formed in the corona or transition region, etc.).
The line of S\,{\sc i} at 1300.907~\AA\ (and of other relatively strong 
S\,{\sc i} lines) shows a behaviour shared by the line at 1318.94~\AA. This
 takes the form of a lower ratio at and to the blue side of line centre, 
without significant emission to the red side of line centre.    

\subsection{Flux ratios in cool stars}        
In the evolved giants that have a larger than solar N/C abundance ratio, the 
N\,{\sc i} lines at 1411.942~\AA\ become apparent (in addition to the 
N\,{\sc i} line at 1319.675~\AA). The ratio of the fluxes in the 1411.942-\AA\
and 1318.94~\AA\ lines becomes substantially larger than in the cool dwarf
 stars. For example, using the fluxes ($F$) given by Ayres et al. (2007), the 
ratio $F$(1411.94~\AA)/$F$(1318.94~\AA) is 1.9 and 1.4 in the giant stars 
$\mu$~Vel and 24~UMa, respectively, but is only 0.2 in the dwarf star 
$\alpha$~Cen~A (HD 128621A). Thus the line at 1318.94~\AA\ does {\it not} 
behave like a line of N\,{\sc i}.
     
Given that the solar spectra obtained with the HRTS and SUMER suggest that the
 line at 1318.94~\AA\ is due to a transition in S\,{\sc i}, the ratio of the 
flux in this line to that of the S\,{\sc i} line at 1300.907~\AA\ has been 
examined in a range of cool stars. 
In the spectra of cool dwarf stars, the ratio $F(1318.94)/F(1300.907)$ is 
0.96~$\pm 0.02$ in $\epsilon$~Eri (HD 22049) (S. Sim, private communication), 
0.98 in $\xi$~Boo~A (HD 131156A) (H. Kay, private communication) and 0.99 in 
$\alpha$~Cen~A (Ayres et al. 2007). In giant stars, the ratio appears to be 
slightly smaller, but accurate measurements of the 1300.907-\AA\ line flux 
are harder because the resonance lines of O\,{\sc i} become broadened by high
 opacities and, when present, the line of Si\,{\sc iii} at 1301.146~\AA\ can 
also be broad. However, the fluxes measured by Ayres et al. (2007) give ratios
 of 0.75 in both $\mu$~Vel and 24~UMa. Using simply the peak intensities (with
 the local background subtracted) in $\iota$~Cap, $\beta$~Cet and $\alpha$~Boo
 (HD 124897) the ratio lies between 0.77 and 0.93. Again, this behaviour 
suggests a similar origin for the lines at 1300.907~\AA\ and 1318.94~\AA.  

The line at 1318.94~\AA\ does not appear to be pumped by radiation in another
 line. For example, its flux ratio relative to the S\,{\sc i} line at 
1295.65~\AA, which is pumped by the O\,{\sc i} 1302.17-\AA\ line, is around 2 
to 3 in dwarf stars, but less than around 0.5 in giant stars where pumping is
 more effective.  

Overall, the solar and stellar observations do not support the identification
of the line at 1318.94~\AA\ with the N\,{\sc i} transition, but do support
the proposal that the line is due to S\,{\sc i}.

\subsection{Other observations} 
Carlsson, Judge \& Wilhelm (1997) used observations with SUMER to study the 
dynamic 
nature of cell interiors in lines of He\,{\sc i} (584.33~\AA), C\,{\sc i} 
(1364.16, 1329.58 and 1156.03~\AA), N\,{\sc i} (1199.55~\AA\ and the 
1318.94-\AA\ line, which they identify with N\,{\sc i}, 1319.00~\AA), 
O\,{\sc i} (1358.51 and 1152.15~\AA), as well as lines of ionized species. In
 the context of the present work, their most interesting results are that the 
response to brightenings in the continuum emission at 1300~\AA\ is slower in 
the 1318.94-\AA\ line than in either C\,{\sc i} (1329.58~\AA) or C\,{\sc ii} 
(1334.53~\AA), and that the line at 1318.94~\AA\ shows a behaviour
that is qualitatively different from that of the N\,{\sc i} line at 
1199.55~\AA.

\section{Possible origins of the line at 1318.94~\AA\ as a S\,{\sc i}
 transition}

\subsection{Laboratory observations of S\,{\sc i} between 1300 and 1320~\AA} 

Fig. 3 gives a partial energy level diagram of S\,{\sc i}, including the
more important lines discussed below. The levels shown above the first 
ionization limit are all from terms in the 3s$^2$3p$^3$($^2$D$^o$)3d or 5s
configurations (in Fig. 3 and hereafter 3d and 5s). The 3d~$^1$F$^o$ level is
currently placed just below the first ionization limit. The 3d~$^1$P$^o$ level
is a known bound state. Energies are taken from Kaufman \& Martin (1993), 
except for the level at 85057~cm$^{-1}$, required to account for the line
at 1318.94~\AA.

\begin{figure}
\resizebox{\hsize}{!}{\includegraphics{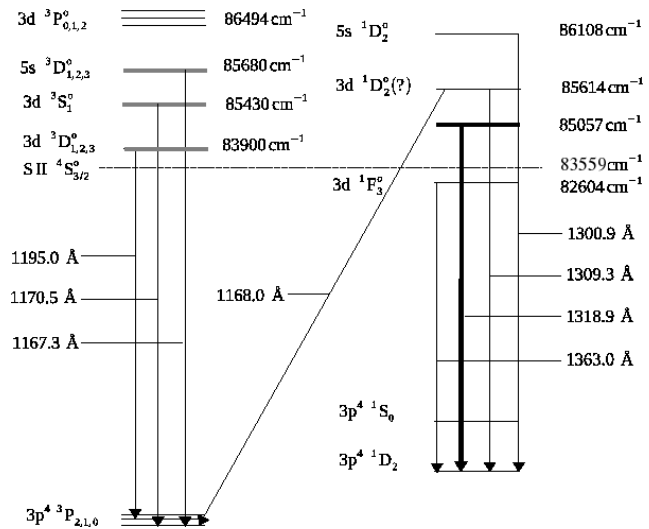}}
\caption{A partial energy level diagram of S\,{\sc i}, showing the energy 
levels and transitions discussed in the text. Not to scale. The light grey
levels are those that can autoionize. The dark grey level is that required to
account for the 1318.94-\AA\ line. The other energies and observed wavelengths
are taken from Kaufman \& Martin (1993).}
\end{figure}

There have been two reports of a line near 1319~\AA\ in laboratory spectra
that were obtained without the presence of lines of N\,{\sc i}, but the
reasons for the absence of the line in other experiments must also be 
considered.

M\"uller (1968) reported unidentified lines of S\,{\sc i} at 1319.0, 1300.0 
and 1228.0~\AA\ in absorption line measurements made using a wall-stabilized
 arc containing SF$_6$. Transitions from all the ground configuration terms of
 S\,{\sc i} were observed and also some lines of S\,{\sc ii}. It is possible 
that the line at 1300.0~\AA\ is the known S\,{\sc i} line at 1300.907~\AA, but
 the wavelengths that M\"uller measured for other lines of S\,{\sc i} agree 
with currently accepted values to within $\pm 0.1$~\AA. Thus the presence of 
the line at 1300.907~\AA\ is not certain. The wavelength of M\"uller's line at
 1319.0~\AA\ is consistent with that observed in the Sun and cool stars at 
1318.94~\AA. Only weak lines of S\,{\sc i} are expected near 1228~\AA. 

M\"uller's source was in equilibrium at a temperature ($T$) of about 
1.2$\times 10^4$~K, and the populations of the levels in the ground 
configuration were expected to be given by the Boltzmann distribution. Since 
the strength of absorption lines depends on both the lower level population 
and the oscillator strength, only the product 
$g_{i}f_{ij} {\rm exp}(-(E_{i}/{\rm {k}} T))$ could be found for the 
unidentified lines. (Here, $g_{i}$ is the statistical weight of the lower 
level, $f_{ij}$ is the absorption oscillator strength and $E_{i}$ is the 
energy of the lower level.) These products were the same for the lines at 
1319~\AA\ and 1300~\AA. If the 1300-\AA\ line {\it is} the transition at 
1300.907~\AA, then the measured value of $g_{i}f_{ij}$ is only 19~per cent 
larger 
than the value (using $f_{\rm length}$) calculated by Deb \& Hibbert (2008) 
and 5~per cent larger than the value from Kurucz (2009). This lends some 
support to the identification of the line with that at 1300.907~\AA. (Other
 calculations of oscillator strengths of S\,{\sc i} lines are discussed by 
Deb \& Hibbert 2008.) In this case, the oscillator strength of the line at 
1318.94~\AA\ would be expected to be very similar to that of the line at 
1300.907~\AA. Overall, the presence of a line of S\,{\sc i} at 1318.94~\AA\ 
in M\"uller's spectrum seems likely, but that of the line at 1300.907~\AA\ is 
not certain. 

Berry et al. (1970) measured the beam-foil emission line spectrum of sulphur
 between 600 and 4000~\AA\ and identified lines of S\,{\sc i} to S\,{\sc vi}. 
Checks were made that showed that carbon, nitrogen and oxygen did not 
contribute to the spectrum. A weak line was reported at 1318~\AA\ and was 
attributed to a low stage of ionization (S\,{\sc i} to S\,{\sc iv}). No line 
was observed at 1301~\AA. Kaufman \& Martin (1993) list a S\,{\sc iv} line at 
1319.04~\AA, so the presence in the source of a line of S\,{\sc i} at 
1318.94~\AA\ is not certain.

Apart from the work by Tondello (1972) and Kaufman (1982, and references 
therein), other experiments have not covered the wavelength region including 
1319~\AA, or have observed only transitions to the ground $^3$P term. 
Tondello (1972) used a flash-pyrolysis system and very pure sulphur powder to 
produce the absorbing medium. Weak lines of C\,{\sc i}, O\,{\sc i} and 
Si\,{\sc i} were also present. He reported and identified the S\,{\sc i} 
lines at 1300.91, 1308.2 and 1309.3~\AA\ for the first time. 

In Tondello's (1972) spectrum, the line at 1308.2~\AA\ is fuzzy (see his fig. 
4) since the upper levels (5s~$^3$D$^{\rm o}$) can autoionize. (The lower 
level is the $^1$D$_2$ level of the ground configuration). The line at 
1309.3~\AA\ is  reported to be broader than other single lines, but might be 
blended with the $^2$P$^{\rm o}_{3/2}$ - $^2$S$_{1/2}$ line of Si\,{\sc ii} 
at 1309.276~\AA, since the weaker member of the Si\,{\sc ii} multiplet seems
 to be present at 1304.370~\AA. There is some doubt about the identity of the 
upper level; in Martin et al. (1990) and Kaufman \& Martin (1993) it is given 
as 3d~$^1$D$^{\rm o}_{2}$, but both the 3d and $^1$D$^{\rm o}_{2}$ assignments
 are accompanied by question marks. The line at 1309.3~\AA\ is identified 
here for the first time in the solar spectrum. It appears in 20~s 
exposures obtained during the HRTS-I rocket flight and is a {\it narrow} line 
that is blended with the Si\,{\sc ii} line at 1309.276~\AA. In Fig. 2 it is 
most obvious near the solar limb, where the Si\,{\sc ii} line (marked 'y') 
becomes optically thick and relatively weaker (see also the line marked 'x', 
which is the weaker member of the Si\,{\sc ii} multiplet). Unlike the 
Si\,{\sc ii} line, the narrow line does not extend above the solar limb. The 
wavelength measured here is 1309.36~$\pm 0.02$~\AA. The average wavelength of 
this line and that of the Si\,{\sc ii} line at 1309.276~\AA\ then agrees with 
Tondello's (1972) listed wavelength of 1309.3~\AA. 

The absence of the 1318.94-\AA\ line from Tondello's (1972) list is a 
potential problem, since he observed absorption lines from all the
terms of the ground configuration, including the lines at 1300.91 and 
1309.3~\AA. As 
discussed in Sections 2.2 and 2.3, in main-sequence stars, the 1300.91-\AA\
 line has a similar intensity (or flux) to that of the 1318.94-\AA\ line.
However, Tondello (1972) states that his table 3 gives new lines that are
observed and classified from his spectra, which are illustrated up to 
1310~\AA. Since no list of observed but unclassified lines is given, the 
presence of a line at 1318.94~\AA\ cannot strictly be excluded. 

Kaufman (1982) reported results from emission line spectra in the wavelength 
range from 1157~\AA\ to 4158~\AA. Only two levels above the first ionization 
limit were included, ($^2$D$^{\rm o}$)5s~$^1$D$^{\rm o}_{2}$ and 
($^2$P$^{\rm o}$)5s~$^1$P$^{\rm o}_{1}$. Lines of N\,{\sc i} were
included in those used for wavelength standards, so if the line at 1318.94~\AA\
were present, it could perhaps have been masked by N\,{\sc i}. 

\subsection{Possible S\,{\sc i} transitions} 
If the line at 1318.94~\AA\ {\it is} due to S\,{\sc i}, what are the possible
upper levels? No previously known levels can account for just one (narrow)
transition to the ground $^3$P term at this wavelength (the upper level would
lie below the $^4$S$^{\rm o}$ ionization limit). A transition to the ground 
configuration $^1$S$_0$ level would require an upper level with ${\rm J} = 1$.
No suitable levels with a $^2$D$^{\rm o}$ or $^2$P$^{\rm o}$ parent are 
expected at the required energy of 97998~cm$^{-1}$. This leaves a decay to the 
$^1$D$_{2}$ level of the ground configuration; the unknown upper level must 
then have $J$ = 1, 2 or 3. With a wavelength of 1318.94~$\pm 0.01$~\AA, the 
energy of the upper level would be 85057.1~$\pm 0.6$~cm$^{-1}$ above the 
S\,{\sc i} ground level, and 1498.0~cm$^{-1}$ above the $^4$S ionization 
limit. The identifications of the autoionizing 3d~$^3$D$^{\rm o}$, 
3d~$^3$S$^{\rm o}$ and 5s~$^3$D$^{\rm o}$ levels seem to be secure. 
The singlet levels are discussed below.

The levels to be considered are (i) 3d~$^1$D$^{\rm o}_2$, currently 
attributed, with question marks, to the level at 85614~cm$^{-1}$, (ii) 
3d~$^3$F$^{\rm o}_{2,3}$ which have so far not been located, and (iii) 
3d~$^1$F$^{\rm o}_3$, currently attributed to a level at 82604.41~cm$^{-1}$ 
(below the $^4$S$^{\rm o}$ ionization limit).

Deb \& Hibbert (2006) carried out theoretical calculations of energy levels 
and oscillator strengths in S\,{\sc i}. They report energy levels for LS 
terms including the effects of configuration interaction (CI) and they 
include levels above the first ionization limit. Fine structure is also 
included in some of their calculations of oscillator strengths. An 
extended list of LSJ oscillator strengths is given by Deb \& Hibbert (2008). 
Although Deb \& Hibbert (2006) state that the accuracy of their calculated
energy levels 
above the $^4$S$^{\rm o}$ ionization limit is not sufficient for individual 
line identifications, there are aspects that are useful to the present work. 

There is a significant difference between the {\it order} of the calculated 
energy levels and those allocated on the basis of laboratory observations. 
In particular, the 3d~$^1$F$^{\rm o}_{3}$ level is calculated to lie
{\it above} the $^4$S$^{\rm o}$ ionization limit, while the 3d~$^3$F$^{\rm o}$
 term is predicted to lie {\it below} this limit. The same order for these 
levels was predicted in the earlier calculations, given in {\sc topbase} (the 
Opacity Project Team 1995). There appears to have been no explicit discussion 
in the literature of the order of the 3d~$^1$F$^{\rm o}$ and 
3d~$^3$F$^{\rm o}$ terms. 

If the order of the calculated levels is assumed to be correct, then the 
assignment of the 5s~$^1$D term and two singlet 3d terms of interest would be 
as follows. The highest level is 5s~$^1$D$^{\rm o}_2$ and can be attributed to
 the shortest wavelength line at 1300.907~\AA. The 3d~$^1$F$^{\rm o}_3$ level 
would be the next highest and would then be identified with the level at 
85614~cm$^{-1}$, instead of 3d~$^1$D$^{\rm o}_2$. The proposed level at 
85057~cm$^{-1}$, would become the 3d~$^1$D$^{\rm o}_2$ level. The 
3d~$^3$F$^{\rm o}$ term would remain unidentified and is calculated to lie 
below the $^4$S$^{\rm o}$ ionization limit. 
 
The evidence for the {\it current} identification of the 3d~$^1$F$^{\rm o}_3$ 
level at 82604.41~cm$^{-1}$ is now considered. The transition to the ground 
configuration $^1$D$_2$ level occurs at 1363.03~\AA\ and according to Kaufman
 \& Martin (1993) has a relatively low intensity in laboratory spectra. Deb 
\& Hibbert (2008) calculate a large transition probability for this 
transition ($A_{\rm length}$ = 4.74 $\times 10^{8}$~s$^{-1}$), having adjusted
their 
calculated energy level to fit the classification by Kaufman (1982). Kurucz 
(2009) has also calculated oscillator strengths for a large number of 
S\,{\sc i} lines, using a semi-empirical approach. His results are
mentioned when they differ significantly from those of Deb \& Hibbert (2008). 
For the line at 1363.03~\AA\ the oscillator strength found by Kurucz 
(2009) is over three orders of magnitude smaller than that calculated by Deb 
\& Hibbert (2008). The line at 1363.03~\AA\ is {\it not} observed in the 
stellar and quiet Sun spectra used in the present studies, although a weak 
line does occur in the solar flare spectrum (Cohen et al. 1978). This is
understandable if the 3d~$^1$F$^{\rm o}_3$ level is populated by collisions 
from the ground configuration $^1$D$_2$ level. The situation in giant stars 
is more complex; the energy of the 3d~$^1$F$_3$ level is close to that of
levels (having the $^4$S$^{\rm o}$ parent) with $n$ = 11, 12 and 13, which can 
be pumped by the H Ly $\alpha$ line (Brown
\& Jordan, 1980). Cascades from these high-$n$ levels lead to relatively 
strong lines of S\,{\sc i} in the {\it uv} spectra of giant stars (Judge 
1988). Thus the current identification of the 3d~$^1$F$^{\rm o}_3$ level with 
that at 82604.41~cm$^{-1}$ depends on one relatively weak line in laboratory 
spectra. No decay to the ground $^3$P$_2$ level has been observed at 
1210.59~\AA, consistent with its much lower transition probability (Deb \& 
Hibbert 2008).  

The level at 85614~cm$^{-1}$ is currently identified as 3d~$^1$D$^{\rm o}_2$
(Kaufman \& Martin 1993), but with question marks (see above). The line at 
1309.3~\AA\ originates from this energy level. Deb \& Hibbert (2008) calculate
 a transition probability ($A_{\rm length}$) that is smaller by a factor 
of 27 than that of the 1300.907-\AA\ line, whereas that found by Kurucz (2009)
is only 10~percent smaller. In the LS calculations of energy levels by 
Deb \& Hibbert (2006), apart the 3d~$^1$F$^{\rm o}_{3}$ level, the 
3d~$^1$D$^{\rm o}_2$ level has the largest difference between the observed 
and calculated values. 

In the Sun and cool dwarfs the 1309.3-\AA\ line is blended with a strong line
 of Si\,{\sc ii} (see Section 3.1) and although distinguishable near the limb
 in solar spectra, its intensity has not been measured. The most one can say 
is that it is a little weaker than, and comparable in width to, the lines at 
1300.907 and 1318.94~\AA. In the K-giants the Si\,{\sc ii} line becomes 
relatively weaker than the S\,{\sc i} line at 1300.907~\AA\ and the total 
flux in the Si\,{\sc ii} line can be used to limit the contribution from the 
S\,{\sc i} 1309.3-\AA\ line, e.g. in $\alpha$~Boo, this limit is $\simeq 0.4$ 
that of the 1300.907-\AA\ line. The expected relative intensity of the 
1309.3 and 1300.907-\AA\ lines does, of course, depend on the excitation 
mechanism, not just the transition probabilities (see Section 4). Also, the 
radiative transfer in the Si\,{\sc ii} and S\,{\sc i} lines needs to be 
treated simultaneously to investigate any pumping of the S\,{\sc i} line 
when the Si\,{\sc ii} line has a high opacity.

Decays from a 3d~$^1$D$^{\rm o}_{2}$ level can also occur to the ground 
$^3$P$_{2,1}$ levels. With the upper level at 85614~cm$^{-1}$, these are 
expected at 1168.03~\AA\ and 1173.46~\AA, respectively. A relatively weak line
line at 1168.03~\AA\ is observed in laboratory spectra 
(Kaufman \& Martin 1993), and at 1168.06~\AA\ in the solar spectrum (Curdt et 
al. 2001). (The latter wavelength appears to originate from Kelly 1987, who
 adopted earlier values for the energy levels of the ground $^3$P term.) No 
line has been reported at 1173.46~\AA. According to the calculations 
by Deb \& Hibbert (2008), both of the above lines have very small branching 
ratios for decays from the upper level and it is surprising that the line at 
1168.03~\AA\ is observed in the Sun. 
 
If instead the level at 85614~cm$^{-1}$ were the 3d~$^1$F$^{\rm o}_3$ level,
the same decays would be expected except that to the $^3$P$_{1}$ level
at 1173.46~\AA. 
   
The possibility that the 3d~$^1$D$^{\rm o}_2$ level lies at 85057~cm$^{-1}$ is
now considered.  If it does, the difference between the energies observed and 
calculated by Deb \& Hibbert (2006) would be reduced from -780~cm$^{-1}$ to
-222~cm$^{-1}$ (more in line with the differences found for other energy 
levels, apart from the current attribution for the 3d~$^1$F$^{\rm o}_3$ level.)
Decays to the ground $^3$P$_{2,1}$ levels would occur at 1175.68~\AA\ and 
 1181.18~\AA. No line at 1181.18~\AA\ has been reported in laboratory 
 spectra; there, a line at 1175.68~\AA\ could be masked by nearby, broad 
autoionizing transitions. In the Sun and cool stars, a line at 1175.68~\AA\ 
would be blended with the strongest member of multiplet uv 4 of C\,{\sc iii}. 
A blend has been indicated in fig. 4 of Curdt et al. (2001) at 
1175.65~\AA, but the possible stage of ionization is not given and the line 
does not appear in their table A.1. No solar line has been reported at 
1181.18~\AA, nor does one appear in the stellar spectra. Thus, without 
knowledge of the transition probabilities, it can be neither confirmed or 
excluded that the proposed level at 85057~cm$^{-1}$ is the 
3d~$^1$D$^{\rm o}_2$ level. 
  
Finally, if some unrecognized interaction reverses the order of the 
3d~$^1$F$^{\rm o}$ and 3d~$^3$F$^{\rm o}$ terms from that calculated, it 
would, in principal, be possible for the level at 85614~cm$^{-1}$ to be the 
$J$ = 2 and/or 3 levels of the 3d~$^3$F$^{\rm o}$ term. This was suggested by 
Joshi et al. (1987), but the higher members of their proposed $n$d-series 
have not been confirmed. However, since the line at 1309.3~\AA\ would be an 
intersystem line, the similar intensities of the lines at 1300.907, 1309.3 
and 1318.94~\AA\ would need to be explained in terms of calculated $A$-values 
and excitation mechanisms. Decays to the ground $^3$P term would also break 
the LS coupling selection rules, but decays at long wavelengths through other 
bound levels are possible.
 
\section{Processes causing the large fluxes in the lines at 1301 and 
1319~\AA}

Judge (1988) gives a full discussion of the excitation mechanisms for 
multiplets of S\,{\sc i} 
observed in spectra of cool giants with the {\it IUE}, and of the ionization 
and recombination mechanisms. In particular, he estimated that photoionization 
dominates over collisional ionization and that low-temperature di-electronic 
recombination is comparable in magnitude to radiative recombination. 
(Low-temperature dielectronic recombination was also found to be a likely 
contributor to the observed flux in multiplets uv 3 and uv 11.) In the 
cool giant stars Judge (1988) estimates that the S\,{\sc i}/S\,{\sc ii} 
population ratio is only about 0.1 at the temperature where the lines are 
formed. In the chromospheres of cool dwarf stars, the electron densities are 
larger, and the relevant radiation fields need to be computed to find the 
above population ratio. 

The populations of the upper levels of the observed solar (or stellar) lines 
of S\,{\sc i} can be found by making use of the observed intensities or 
fluxes, a solar or stellar model and the transition probabilities. Full 
radiative transfer also needs to be included for lines and continua. Here, 
the aim is to simply identify the processes that lead to the relatively 
strong lines at 1300.907 and 1318.94~\AA. The autoionizing triplet levels are 
considered first.  

Wilhelm et al. (2005) discovered broad emission features in the solar spectrum 
around 1170~\AA, which Avrett, Kurucz \& Loeser (2006) identified as radiative
 decays of the S\,{\sc i} autoionizing 3d~$^3$S$^{\rm o}_{1}$ and 
5s~$^3$D$^{\rm o}_{1,2,3}$ levels, to the levels of the ground $^3$P term. The
 strongest features lie at 1170.55 and 1167.13~\AA, respectively, adopting the
 calculated wavelengths from Kaufman \& Martin (1993). Using the widths 
measured by Tondello (1972) leads to autoionization rates of 
$\simeq 10^{13}$~s$^{-1}$. The radiative decay rates calculated by Deb \& 
Hibbert (2008) are far smaller, $\le 3 \times 10^8$~s$^{-1}$. Under these 
circumstances (provided rates to and from other levels are much smaller than 
the autoionization rate), the ratio, 
$n_{\rm u}/(n_{\rm S\,\hbox{\tiny II}}n_{\rm e})$ 
is determined by detailed balance between autoionization and dielectronic 
capture. Here, $n_{\rm u}$, $n_{{\rm S\,\hbox{\tiny II}}}$ and $n_{\rm e}$ 
are the number densities of the upper level, the ground state of  S\,{\sc ii} 
and the electrons, respectively. The S\,{\sc i} lines from the autoionizing 
levels are therefore low temperature equivalents of the high temperature 
dielectronic satellite lines found in the He\,{\sc i}-like ions (Gabriel \& 
Jordan 1969). (The relevant theory is set out in Gabriel \& Paget 1972). 
The above ratio is then given by the Saha -- Boltzmann equation. Using the 
statistical weights of the 5s~$^3$D$^{\rm o}$ autoionizing levels (level u) 
and the $^4$S$^{\rm o}_{3/2}$ ground state of S\,{\sc ii}, and the energy of 
the 5s~$^3$D$^{\rm o}$ term above the ionization limit, leads to
\begin{equation}
\frac{n_{\rm u}}{n_{{\rm S\,\hbox{\tiny II}}}} = 7.8 \times 10^{-16} 
               n_{\rm e} T_{\rm e}^{-3/2} 10^{(-1325/T_{\rm e})}~.  
\end{equation}
Thus the ratio $n_{\rm u}/n_{\rm S\,\hbox{\tiny II}}$ can be found as a 
function of $T_{\rm e}$, provided $n_{\rm e}$ is known from a solar (or 
stellar) model. 

Avrett et al. (2006) made a preliminary model that accounted for the strength
of the autoionizing lines observed in the SUMER spectra, but they stressed 
that a more detailed model was being prepared. This has been published by 
Avrett \& Loeser (2008) and is based on lines of hydrogen, carbon and oxygen. 
Avrett et al. (2006) also illustrate the contribution function for the 
1167.13-\AA\ line, which peaks at about 5720~K, but the line formation 
process was not discussed. Over the temperature range where the 1167.13-\AA\ 
line is mainly formed in the Avrett et al. (2006) model, the values of
$n_{\rm u}/n_{\rm S\,\hbox{\tiny II}}$ in the Avrett \& Loeser (2008) model 
are larger by only about a factor 1.5, this factor reflecting 
larger values of $n_{\rm e}$. The values of the ratio at 5720~K are 
$\simeq 8 \times 10^{-11}$ and $\simeq 1.2\times 10^{-10}$, respectively.
 
The value of $n_{\rm u}$ can be expressed as
\begin{equation}
n_{\rm u} = \frac{n_{\rm u}}{n_{\rm S\,\hbox{\tiny II}}}
            \frac{n_{\rm S\,\hbox{\tiny II}}}{n_{\rm S}}
            \frac{n_{\rm S}}{n_{\rm H}} n_{\rm H}~.     
\end{equation}
Using a sulphur abundance of 2 $\times 10^{-5}$ 
that of hydrogen and $n_{\rm S\,\hbox{\tiny II}}/n_{\rm S} \le 1.0$ gives
an upper limit for $n_{\rm u}$ as a function of $T_{\rm e}$. At 5720~K, the
 values are 0.15 and 0.035~cm$^{-3}$, using the values of $n_{\rm H}$ from 
the models by Avrett \& Loeser (2008) and Avrett et al. (2006), respectively.
    
The intensity of the decay from $n_{\rm u}$, for an optically thin line in a 
plane parallel atmosphere, is given by 
\begin{equation}
I = \frac{{\rm h} {\rm c}}{4 \pi \lambda} \int n_{\rm u} A_{\rm r} {\rm d}h, 
\end{equation}
where $A_{\rm r}$ is the radiative decay rate from the upper level, u, and 
${\rm d}h$ is the path length through the atmosphere. The lines of interest 
show brightening as the solar limb is approached (see Avrett et al. 2006), so 
they are unlikely to have a very high opacity. To allow for some opacity, the 
constant of 4$\pi$ in equation (1) is replaced by 2$\pi$ in the calculations 
that follow. This introduces an uncertainty of a factor of two in the upper 
level populations, since with very high opacities, the constant would be $\pi$
 (see discussion in Pietarila \& Judge 2004). 

Using the observed peak radiance (Curdt et al. 2001) with the continuum
 removed, the line width from Tondello (1972) and the value of $A_{\rm r}$
from Deb \& Hibbert (2008), the value of $\int n_{\rm u} {\rm d}h$ can be 
found and is 2.3 $\times 10^5$~cm$^{-2}$. Using the half-width of 
the extent of the emitting region from the contribution function for the 
1167.13-\AA\ line given in Avrett et al. (2006) (this is not available for the
model by Avrett \& Loeser 2008), the mean value of $n_{u}$ for the 
5s~$^3$D$^{\rm o}$ levels can be found. This is 
$\simeq 1.4\times 10^{-2}$~cm$^{-3}$, consistent with the upper limit of 
3.5$\times 10^{-2}$~cm$^{-3}$ given above. The fact that the estimated 
population density of the 5s~$^3$D$^{\rm o}$ term is only a factor of 2.5 
lower than the upper limit expected from the Saha -- Boltzmann equation, when 
$n_{\rm S\,\hbox{\tiny II}}/n_{\rm S}$ = 1 is adopted, strongly supports
the proposal that the autoionizing levels are populated by low-temperature
dielectronic capture. 

Ignoring any decays from higher levels, the dielectronic recombination rate 
coefficient through the 5s~$^{3}$D$^{\rm o}$ levels can also be found, by 
taking into account the fraction of the dielectronic captures that undergo 
radiative decay. The value is $\simeq 3 \times 10^{-14}$~cm$^{3}$~s$^{-1}$, 
using the $A$-values of Deb \& Hibbert (2008). Judge (1988) estimated the 
{\it total} low-temperature dielectronic recombination rate coefficient 
through levels within 5,000~cm$^{-1}$ of the S\,{\sc ii} ground state to be 
$\simeq 10^{-13}$~cm$^{3}$~s$^{-1}$, using the method set out by Storey (1981)
 (see also Nussbaumer \& Storey 1983), although only approximate radiative 
rates were then available.

The relative intensity of the 1300.907-\AA\ line to that of the 1167.13-\AA\ 
line can be used to find the population of the 5s~$^{1}$D$^{\rm o}_{2}$ 
level relative to that of 5s~$^3$D$^{\rm o}$, making use of the $A$-value by
Deb \& Hibbert (2008) for the 1300.907-\AA\ transition. The intensity of the 
1300.907-\AA\ line is found from the radiance given by Curdt et al. (2001) 
with the local continuum removed. It is assumed that this line is formed over 
the same region of the solar atmosphere as the 1167.13-\AA\ line. This 
approach leads to a population of the 5s~$^{1}$D$^{\rm o}_{2}$ level that is 
about 0.1 times that of the 5s~$^3$D$^{\rm o}$ term. 

Even when it is assumed that all sulphur is in the form of S\,{\sc i}, 
and with generous estimates for the collisional excitation rate from 
3p$^4$~$^1$D$_2$, and for $T_{\rm e}$ and $n_{\rm e}$, it is
clear that the 5s~$^1$D$^{\rm o}_2$ level population is several orders of 
magnitude too large to be accounted for by collisions from the ground 
configuration $^1$D$_2$ level (see also Judge 1988). 

The large population of the 5s~$^{1}$D$^{\rm o}_2$ level is not surprising 
since the wavenumber region between 83900~cm$^{-1}$ and 
85680~cm$^{-1}$ contains three autoionizing terms that are expected to be
heavily populated by di-electronic capture. This region also contains
the proposed level at 85057~cm$^{-1}$, required to account for the line at 
1318.94~\AA. It is therefore possible that the non-autoionizing levels
in this region are populated mainly by collisions between these and the
autoionizing levels. With the small energy intervals involved, collisions with
heavy particles as well as electrons need to be considered. Also, since the
autoionization rates are so much larger than the radiative decay rates, only 
a small amount of spin-orbit mixing between the autoionizing 
5s~$^3$D$^{\rm o}_{2}$ level and the 5s~$^1$D$^{\rm o}_{2}$ level, and
 between the
autoionizing 3d~$^3$D$^{\rm o}_{2}$ level and the 3d~$^1$D$^{\rm o}_{2}$ level 
could lead to dielectronic capture directly into these singlet states. Using 
the non-thermal width of the 1300.907-\AA\ line in STIS spectra of 
$\epsilon$~Eri (Sim \& Jordan 2003) the
autoionizing rate involved must be less than $\simeq 10^{11}$~s$^{-1}$, 
 but could still be significantly larger than the radiative rate of 
2.9$\times 10^8$~s$^{-1}$ (Deb \& Hibbert 2008). In either case, it is the
 presence of the strongly autoionizing states in S\,{\sc i} that leads to the
 strength of the decays from the singlet levels.

\section{The 1309 -- 1310~\AA\  emission feature in cool giants}

In cool giants, the spectrum between 1309 and 1310~\AA\ becomes more complex
than in the cool dwarf stars. The Si\,{\sc ii} line at 1304.370~\AA\ becomes
blended with the blue wing of the O\,{\sc i} line at 1304.858~\AA. Pumping by
this O\,{\sc i} line distorts the normal profile of the Si\,{\sc ii} line at
1309.275~\AA\ (Jordan \& Judge 1984; Munday 1990). Examples of this can be 
seen in the G-type giants 24~Uma and HR~9024. In cooler giants such as 
$\mu$~Vel, $\iota$~Cap and $\beta$~Cet, another distinct line becomes apparent
 at $\simeq$ 1310~\AA\ and this becomes the dominant feature in the K-giants
$\alpha$~Boo and $\alpha$~Tau. The additional presence of the S\,{\sc i} line 
at 1309.36~\AA\ is now known to complicate the spectrum. Although, from 
spectra obtained with {\it IUE}, Jordan \& Judge (1984) thought that much of 
the feature in $\alpha$~Tau might be due to Si\,{\sc ii}, the presence of the 
separate line in the STIS spectra of the above giants suggests that this is 
not the case. A possible contribution from a line of P\,{\sc ii} was 
discussed, but was thought to be unlikely given the low abundance of 
phosphorus. This line has a wavelength of 1309.87~\AA, consistent with that 
measured here from the STIS spectra, and is the 
3s$^{2}$3p${^2}$~$^3$P$_2$ -- 3s3p$^3$~$^3$P$^{\rm o}_1$ transition. The upper
 level can be pumped from the lower $^3$P$_1$ level at 1304.68~\AA, by the 
O\,{\sc i} 1304.858~\AA\ line. There is also a less close coincidence between 
the P\,{\sc ii} $^3$P$_0$ -- $^3$P$^{\rm o}_1$ line and the O\,{\sc i} line at
 1302.168~\AA. The wavelength coincidence with the more effective O\,{\sc i} 
pumping line is closer for the P\,{\sc ii} line than for the Si\,{\sc ii} 
line, so the higher O\,{\sc i} line opacities in the K-giants could 
counteract the lower abundance of phosphorus to produce an observable line. 
In cool stars, a relatively strong line of Cl\,{\sc i} at 1351.66~\AA\ 
(chlorine has a similar abundance to that of phosphorus) is produced by 
photoexcitation by the C\,{\sc ii} resonance line at 1335.71~\AA\ (Shine 
1983). 
       
\section{Discussion and conclusions}

The observational evidence from both solar and stellar {\it uv} spectra 
strongly suggests that the line near 1319~\AA\ is {\it not} due to N\,{\sc i},
as commonly adopted in the literature. The observed wavelength of 
1318.94 $\pm 0.01$~\AA, measured relative to 
other lines of neutral atoms, does not agree with that of the N\,{\sc i} line
at 1319.001~\AA\ and the intensity (or flux) in the observed line is much 
larger than
that expected from the theoretically stronger member of the multiplet at 
1319.675~\AA. Also, between giant stars where the nitrogen abundance is 
enhanced and dwarf stars, the ratio of the N\,{\sc i} lines at 1411.942~\AA\ 
to that of the 1318.94-\AA\ line becomes significantly larger. The wavelength
calibration of spectroscopic instruments that have assumed that the line at 
1318.94~\AA\ is the N\,{\sc i} line at 1319.001~\AA\ need to be reexamined in 
this wavelength region.     

It is proposed that the unidentified line at 1318.94~\AA\ is due to a 
transition in S\,{\sc i} from above the $^4$S$^{\rm o}$ ionization limit.  
On the basis of the {\it order} of the energy levels calculated by Deb \&
Hibbert (2006), the proposed level at 85057~cm$^{-1}$ is identified with
the 3d~$^{1}$D$^{\rm o}_2$ level. This was previously tentatively assigned to 
a level at 85614~cm$^{-1}$ (Martin et al. 1990; Kaufman \& Martin 1993). This
 identification reduces the 
difference between the observed energy level and that calculated by Deb \& 
Hibbert (2006). The wavelengths of the possible decays to the ground 
$^3$P$_{2,1}$ levels are 1175.68 and 1181.18~\AA. The latter line is not 
observed and blending prevents the detection of the line at 1175.68~\AA.

The level at 85614~cm$^{-1}$ could then be 3d~$^1$F$^{\rm o}_3$. However, if 
the order of the 3d~$^1$F$^{\rm o}$ and 3d~$^3$F$^{\rm o}$ terms is reversed 
compared with that calculated, an alternative classification of the level at 
85614~cm$^{-1}$ would be one or both of the 3d~$^3$F$^{\rm o}_{3,2}$ levels, 
as suggested by Joshi et al. (1987). The expected weakness of confirming 
decays to the ground $^3$P term prevents a definite conclusion. The decay to
the ground configuration $^1$D$_2$ level at 1309.36~\AA\ has been observed 
for the first time in an astrophysical spectrum (in the Sun). In contrast to 
the conclusion by Tondello (1972), this line
does not appear to be significantly broader than that at 1300.907~\AA, 
although blending with the Si\,{\sc ii} line at 1309.276~\AA\ prevents a
proper measurement.
 
One test of the proposed identifications would be to use them to 'fine tune' 
the calculated energy levels that are adopted in the calculations of the 
oscillator strengths, in the manner described by Deb \& Hibbert (2006). 
For example, the $A$-value of the line at 1309.36~\AA\ might be 
smaller if the upper level were part of the 3d~$^3$F$^{\rm o}$ term than if it
 were the 3d~$^1$F$^{\rm o}_3$ level. The resulting relative line intensities 
could then be tested against those observed. All the lines of S\,{\sc i} 
discussed here need to be included in calculations of model atmospheres. 
They can provide good tests of chromospheric electron densities and in-situ 
radiation fields because both control the relative population densities of 
S\,{\sc i} and S\,{\sc ii}.   

The overall strength of the observed lines from the levels in the 
($^2$D$^{\rm o}$)3d and 5s configurations can be attributed to low-temperature
dielectronic recombination. In the Sun, the upper level populations of the 
lines examined are within about an order of magnitude of those given by the 
Saha -- Boltzmann equation.

As part of the work on the region around the Si\,{\sc ii} line at 1309.275~\AA,
a distinct line at 1309.87~\AA\ that becomes apparent in the late-G/early-K 
giants has been identified as a P\,{\sc ii} transition, pumped by the strong 
O\,{\sc i} line at 1304.86~\AA. This becomes much stronger than the 
Si\,{\sc ii} line in the cooler K-giants $\alpha$~Boo and $\alpha$~Tau.  \\
~\\
{\bf ACKNOWLEDGMENTS}\\
\noindent 
I am grateful to Rachel Koncewicz and the Media and Services group (Oxford
 Physics Department) for producing the figures, to Prof. Hibbert for 
drawing my attention to the paper by Deb \& Hibbert (2006) and to Prof. M.
Carlsson for early discussions about the solar properties of the line near
1319~\AA. This research has made use of the Multimission Archive at the Space 
Telescope Science Institute (MAST). \\
~\\
{\bf REFERENCES} \\
~\\
Avrett E. H., Kurucz R. L., Loeser R., 2006, A\&A, \\
\indent 452, 651 \\
\noindent
Avrett E. H., Loeser R., 2008, ApJS, 125, 229 \\
\noindent
Ayres T. R., Hodges-Kluck E., Brown A., 2007, ApJS, 171, \\
\indent 304 \\
\noindent 
Berry H. G., Schectman R. M., Martinsson I., Bickel W. S., \\
\indent Bashkin S., 1970, JOSA, 60, 335 \\
\noindent
Brown A., Jordan C., 1980, MNRAS, 191, 37P \\
\noindent
Burton W. M., Ridgeley A., 1970, Sol. Phys., 14, 3 \\
\noindent
Carlsson M., Judge P.G., Wilhelm K., 1997, ApJ, 486, L63 \\
\noindent
Chipman E., Bruner E. C., 1975, ApJ, 200, 765 \\
\noindent
Cohen L., Feldman U., Doschek G. A., 1978, ApJS, 37, 393 \\
\noindent
Curdt W., Brekke P., Feldman U., Wilhelm K., Dwivedi B. \\
\indent N., Sch\"ule U., Lemaire P., 2001, A\&A, 375, 591 \\
\noindent
Deb N. C., Hibbert A., 2006, J. Phys. B., 39, 4301 \\
\noindent
Deb N. C., Hibbert A., 2008, ADNDT, 94, 561  \\
\noindent
Gabriel A. H., Jordan C., 1969, Nat, 221, 947 \\
\noindent
Gabriel A. H., Paget T. M., 1972, J.Phys.B, 5, 673 \\
\noindent
Jordan C., 1988, JOSA B, Vol.5, No.10, 2252 \\ 
\noindent
Jordan C., Judge P. G., 1984, Phys. Scripta, Vol.T8, 43      \\
\noindent
Jordan C., Brueckner G. E., Bartoe J.-D. F., Sandlin G. \\
\indent D., VanHoosier M. E., 1978, ApJ, 226, 687 \\  
\noindent
Joshi Y. N., Mazzoni M., Nencioni A., Parkinson W. H., \\
\indent Cantu A., 1987, J. Phys. B, 20, 1203 \\
\noindent
Judge P. G., 1988, MNRAS, 231, 419        \\
\noindent
Kaufman V., 1982, Phys. Scripta, 26, 439 \\
\noindent 
Kaufman V., Martin, W. C., 1993, J. Phys. Chem. Ref. \\
\indent Data, 22(2), 279 \\
\noindent
Kelly R. L., 1987, J. Phys. Chem. Ref. Data, 16, 1 \\
\noindent  
Kurucz R. L., 2009, http://kurucz.harvard.edu/atoms/ \\
\indent 1600 \\
\noindent
Kurucz R. L., Peytremann E., 1975, SAO Special Report \\
\indent No. 362, Smithsonian Institution and Astrophysical \\
\indent Observatory, Cambridge, Mass. \\
\noindent
Martin W. C., Zalubas R., Musgrove A., 1990, J. Phys. \\
\indent Chem. Ref. Data, 19(4), 821 \\
\noindent
Moore C. E., 1975, Nat. Stand. Ref. Data Ser., NBS, 3, \\
\indent Section. 5. \\
\noindent
M\"uller D., 1968, Zeit. Naturforsch, 23a, 1707 \\
\noindent 
Munday M. 1990, D.Phil. Thesis, University of Oxford \\
\noindent
Nussbaumer H., Storey P. J., 1983, A\&A, 126, 75 \\
\noindent
Pietarila A., Judge P. G., 2004, ApJ, 6060, 1239 \\
\noindent
Sandlin G. D., Bartoe J.-D. F., Brueckner G. E., Tousey \\
\indent R., VanHoosier M. E., 1986, ApJS, 61, 801 \\  
\noindent
Shine R. A., 1983, ApJ, 266, 882 \\
\noindent 
Sim S. A., Jordan C., 2003, MNRAS, 341, 517 \\ 
\noindent
Storey P. J., 1981, MNRAS, 195, 27P \\
\noindent
The Opacity Project Team, 1995, The Opacity Project, \\
\indent Vol. 1, IoP Publications, Bristol, UK \\
\noindent
Tondello G., 1972, ApJ, 172, 771 \\
\noindent
Wilhelm K., Sch\"ule U., Curdt W., Hilchenbach M., Marsch \\
\indent E., Lemaire P., Bertaux J.-L., Jordan S. D., Feldman \\
\indent U., 2005, A\&A, 439, 701 

\label{lastpage}

\end{document}